\documentclass{elsart} 
\usepackage{array}
\usepackage{oldlfont}
\usepackage{amssymb}
\usepackage{amsmath}
\journal{Physics Letters A}
\begin{document}
\begin{frontmatter}
\title{On the decrease of the number of bound states with the increase of the angular momentum}
\author{Fabian Brau\thanksref{fnrs}}
\thanks[fnrs]{FNRS Postdoctoral Researcher}
\address{Service de Physique G\'en\'erale et de Physique des Particules El\'ementaires, Groupe de Physique Nucl\'eaire Th\'eorique, Universit\'e de Mons-Hainaut, Mons, Belgique}
\ead{fabian.brau@umh.ac.be}
\date{\today}

\begin{abstract}
For the class of central potentials possessing a finite number of bound states and for which the second derivative of $r V(r)$ is negative, we prove, using the supersymmetric quantum mechanics formalism, that an increase of the angular momentum $\ell$ by one unit yields a decrease of the number of bound states of at least one unit: $N_{\ell+1}\le N_{\ell}-1$. This property is used to obtain, for this class of potential, an upper limit on the total number of bound states which significantly improves previously known results.
\end{abstract}

\end{frontmatter}

\section{Introduction}
\label{sec1}
It is well known that the addition of a positive contribution to a potential lying a finite number bound states can only lessen this number. In the case of central potentials, this property implies that the number of $\ell$-wave bound states is always greater or equal to the number of $(\ell+1)$-wave bound states, $N_{\ell+1} \le N_{\ell}$, $\ell$ being obviously the angular momentum. An accurate knowledge of the variation of the number of $\ell$-wave bound states with the variation of the angular momentum is important to derive cogent upper or lower limits on the {\it total} number of bound states. Indeed, it has been shown in Ref. \cite{br03b} that such limits, contrary to limits on the number of S-wave or $\ell$-wave bound states, yield in general poor results.

A first information about this variation of the number of bound states with the angular momentum can be found from the Bargmann-Schwinger upper limit \cite{ba52,sc61}
\begin{equation}
\label{eq1}
N_{\ell}\le \frac{1}{2\ell+1}\int_0^{\infty}dr\, r\, |V^-(r)|,
\end{equation} 
where $V^-(r)$ is the negative part of the potential obtained by setting its positive part to zero. Unfortunately, this upper limit is not accurate for potentials possessing many bound states since it is proportional to the strength of the potential while the correct evolution is proportional to the square-root of this strength (see for example \cite{ca65a,ca65b}). This implies that the variation of $N_{\ell}$ with respect to $\ell$, as derived from the above constraint, is also not very accurate. Indeed, the relation (\ref{eq1}) inform us that for $\ell$ going from 0 to 1, the number of bound states decreases by a factor 3. For most potentials, the decrease is almost constant and is about one unit.

A more reliable information can be found from the generalization of the Calogero-Cohn upper limit \cite{ca65a,co65} to nonzero angular momentum (which is applicable only to monotonic potential) \cite{ch95}
\begin{subequations}
\label{eq2}
\begin{equation}
\label{eq2a}
N_{\ell}\leq \frac{2}{\pi}\int_0^{\infty}dr\, |V(r)|^{1/2}+1-\sqrt{1+(2/\pi)^2 \,\ell(\ell+1)}, 
\end{equation} 
\begin{equation}
\label{eq2b}
N_{\ell}< \frac{2}{\pi}\int_0^{\infty}dr\, |V(r)|^{1/2}-\frac{2}{\pi}\, \ell+1-\frac{1}{\pi}.
\end{equation} 
\end{subequations}
The upper limit (\ref{eq2}) yields stronger restrictions on $N_{\ell}$, when the potential is strong enough to bind several bound states, than the Bargmann-Schwinger upper limit thanks to the correct behavior of this limit with the strength of the potential. The less stringent relation (\ref{eq2b}) is written to exhibit the linear behavior of this upper limit with a variation of $\ell$. 

Another linear behavior was found in Ref. \cite{br03b} with the following lower limit
\begin{subequations}
\label{eq2c}
\begin{equation}
\label{eq2ca}
N_{\ell}> \frac{1}{\pi}\int_0^{\infty}dr\, |V(r)|^{1/2}-\frac{1}{4\pi}\ln\left[\frac{V(p)}{V(q)}\right]-\frac{\ell}{\pi} \ln\left(\frac{q}{p}\right)-\frac{3}{2}, 
\end{equation} 
where $p$ and $q$ are defined by
\begin{equation}
\label{eq2cb}
\int_0^{p}dr\, |V(r)|^{1/2}=\frac{\pi}{2} \quad \text{and} \quad \int_q^{\infty}dr\, |V(r)|^{1/2}=\frac{\pi}{2}. 
\end{equation}
\end{subequations} 
This lower limit is written for monotonic potentials but a generalization exist \cite{br03b}.

We believe that these kind of variations of the number of $\ell$-wave bound states with the angular momentum $\ell$, illustrated by (\ref{eq2}) and (\ref{eq2c}), are rather close to the exact law of variation, contrary to the variation exhibit by (\ref{eq1}). Indeed, it is known that when the strength, $g$ ($V(r)=g\, v(r)$), of the potential goes to infinity we have \cite{ch68}
\begin{equation}
\label{eq3}
\lim_{g\rightarrow \infty} N_{\ell}(g\, v)\left[\frac{1}{\pi} \int_0^{\infty}dr\, [g\, v^-(r)]^{1/2}\right]^{-1}=1.
\end{equation} 
This means that asymptotically we have $N_{\ell}(g\, v)\approx \alpha g^{1/2}$, where $\approx$ means asymptotic equality and where $\alpha=(1/\pi)\int_0^{\infty}dr\, [v^-(r)]^{1/2}$ is a constant (which eventually depends on various parameters of the potential, except the strength). Suppose now that we have an upper bound on $N_0$ which is proportional to $g^{1/2}$:
\begin{equation}
\label{eq4}
N_0\le g^{1/2}\, {\cal F}[v^-(r)]\equiv N^+_0,
\end{equation} 
where ${\cal F}$ is an operator acting on $v^-(r)$ to give a number. The Calogero-Cohn \cite{ca65a,co65} and the Martin upper limits \cite{ma77} are two examples. A generalization of this upper limit (\ref{eq4}) to nonzero angular momentum cannot be of the form
\begin{equation}
\label{eq5}
N_{\ell}\le \frac{g^{1/2}}{C_{\ell}}\, {\cal F}[v^-(r)],
\end{equation} 
with $C_{\ell}\rightarrow \infty$ as $\ell \rightarrow \infty$ (because for any values of $g$, if $\ell$ is large enough, there is no bound states). Indeed, for $\ell$ large enough, the number ${\cal F}[v^-(r)]/C_{\ell}$ will be smaller than $\alpha$ (see above) which contradicts (\ref{eq3}). A possible generalization will be then of the form
\begin{equation}
\label{eq6}
N_{\ell}\le g^{1/2}\, {\cal F}[v^-(r)]+D_{\ell}\equiv N^+_0+D_{\ell},
\end{equation} 
with $D_{\ell}\le 0$. Obviously, this simple analysis is not sufficient to determine if the variation of the number of $\ell$-wave bound states with $\ell$ is linear or not. Nevertheless, we believe that a linear variation, with eventually a coefficient which depend on the potential (see for example (\ref{eq2c})), is a very good approximation of the correct evolution of $N_{\ell}$ with the angular momentum $\ell$.

\section{Supersymmetry and variation of $N_{\ell}$ with $\ell$}
\label{sec2}

We show in this section how supersymmetry in quantum mechanics (see for example \cite{wi81,su85a,su85b,co95}) can be used to obtain information about the decrease of the number of $\ell$-wave bound states with an increase of the angular momentum.

Let $U^{(\ell)}_0(r)$ be an effective $\ell$-wave potential possessing a finite number, $N_{\ell}$, of $\ell$-wave bound states
\begin{equation}
\label{eq7}
U^{(\ell)}_0(r)=V(r)+\frac{\ell(\ell+1)}{r^2}.
\end{equation}
The superpotential, $W(r)$, associated to $U^{(\ell)}_0(r)$ is defined by the Riccati equation
\begin{equation}
\label{eq8}
W'(r)+ W^2(r)=U^{(\ell)}_0(r)-E^{(\ell)}_0,
\end{equation}
where appended prime means of course differentiation with respect to the radius $r$ and where $E^{(\ell)}_0$ is the energy of the ground state of $U^{(\ell)}_0(r)$. We use the notation $W(r)$ instead of $W^{(\ell)}(r)$ for simplicity. The supersymmetric partner $U^{(\ell)}_1(r)$ of $U^{(\ell)}_0(r)$ is obtained from $U^{(\ell)}_0(r)$ and $W(r)$ with the following relation
\begin{equation}
\label{eq9}
U^{(\ell)}_1(r)=U^{(\ell)}_0(r)-2W'(r).
\end{equation}
It can be shown that $U^{(\ell)}_0(r)$ and $U^{(\ell)}_1(r)$ share the same spectrum except for the ground state which is missing for $U^{(\ell)}_1(r)$ \cite{su85a,su85b,co95} (this remarkable property has been previously used several times to clarify some results in quantum mechanics, see for example \cite{baye87,gang94}). The potential $U^{(\ell)}_1(r)$ has then one energy level less than $U^{(\ell)}_0(r)$. From (\ref{eq8}), it is easy to see that the superpotential $W(r)$ is linked to the wave function of the ground state of $U^{(\ell)}_0(r)$, noted $u^{(\ell)}_0(r)$, by the relation
\begin{equation}
\label{eq10}
W(r)=\frac{d}{dr} \ln u^{(\ell)}_0(r).
\end{equation}
It is well known that the wave function $u^{(\ell)}_0(r)$ behaves as $r^{\ell+1}$ near the origin. This implies that $U^{(\ell)}_1(r)$ has the following behavior for small values of $r$
\begin{equation}
\label{eq11}
U^{(\ell)}_1(r)\sim U^{(\ell)}_0(r)+\frac{2(\ell+1)}{r^2}\sim \frac{(\ell+1)(\ell+2)}{r^2}.
\end{equation}
More precisely, we can write
\begin{equation}
\label{eq12}
U^{(\ell)}_1(r)=V(r)+\frac{(\ell+1)(\ell+2)}{r^2}-2\frac{d^2}{dr^2}\ln\left(\frac{u^{(\ell)}_0(r)}{r^{\ell+1}}\right).
\end{equation}
The supersymmetric partner $U^{(\ell)}_1(r)$ of $U^{(\ell)}_0(r)$ can thus be considered as an effective potential with an angular momentum equal to $\ell+1$. Suppose now that
\begin{equation}
\label{eq13}
\frac{d^2}{dr^2}\ln\left(\frac{u^{(\ell)}_0(r)}{r^{\ell+1}}\right)\ge 0,
\end{equation}
for all values of $r$. The comparison between the definition (\ref{eq7}) of $U^{(\ell)}_0(r)$ and the relation (\ref{eq12}) yields the inequality
\begin{equation}
\label{eq14}
U^{(\ell)}_1(r)\leq U^{(\ell+1)}_0(r),
\end{equation}
which implies, with standard comparison theorem, that the effective potential $U^{(\ell+1)}_0(r)$ has no more bound states than $U^{(\ell)}_1(r)$. Since $U^{(\ell)}_1(r)$ has $N_{\ell}-1$ bound states this entails that $U^{(\ell+1)}_0(r)$ has at most $N_{\ell}-1$ bound states.

Now, we need to find the class of potentials for which the condition (\ref{eq13}) is true. This problem has been treated previously and it was proved (three times!) that the class we search is composed of potentials for which the Laplacian is negative for all values of $r$ \cite{ba84,as88,ma90}:
\begin{equation}
\label{eq15}
\Delta V(r) \leq 0 \quad \text{for} \quad 0\leq r < \infty.
\end{equation}
This condition is obviously equivalent to the requirement that the second derivative of $r V(r)$ be negative. The potentials which enter in the class defined by the inequality (\ref{eq15}) loose at least one bound state when the angular momentum $\ell$ increases by one unit. The Yukawa potential is a simple example.

\section{Upper limit on the total number of bound states}
\label{sec3}

In the previous Section \ref{sec2}, we have identified a class of potentials for which the number of $\ell$-wave bound states is related to the number of $(\ell+1)$-wave bound states by the inequality 
\begin{equation}
\label{eq23}
N_{\ell+1}\leq N_{\ell}-1,
\end{equation}
which entails that
\begin{equation}
\label{eq24}
N_{\ell}\leq N_0-\ell.
\end{equation}
Let $N^+_0$ be any upper limit on $N_0$ ($N^+_0\le N_0$) and let $L^+$ be an upper limit on $L$, the largest value of $\ell$ for which bound states do exist, entailing that for $\ell>L$ the potential certainly does not possess any $\ell$-wave bound state. An upper limit on the total number of bound states is then given by
\begin{equation}
\label{eq25}
N\leq \sum_{\ell=0}^{L} (2\ell+1) (N_0-\ell)\leq \sum_{\ell=0}^{L^+} (2\ell+1) (N^+_0-\ell).
\end{equation}
A simple calculation yields
\begin{equation}
\label{eq26}
N\leq N^+_0 (L^+ +1)^2-\frac{1}{6}\,L^+ (L^+ +1)(4L^+ +5).
\end{equation}

Now, we need to choose suitable upper limits $N^+_0$ and $L^+$. A very stringent upper limit on $N_0$, especially for strong potential possessing many bound states, have been found in Ref. \cite{br03a} for monotonically increasing potentials and generalized in Ref. \cite{br03b} to non-monotonically increasing potentials. To illustrate the improvement yielded by the relation (\ref{eq26}) over existing upper limits, we consider, for simplicity, only monotonic potentials (thus these potentials are purely negative). The upper limit on $N_0$ found in Ref. \cite{br03a} reads
\begin{subequations}
\label{eq27}
\begin{equation}
\label{eq27a}
N_0\leq \frac{1}{\pi}\int_0^{\infty}dr\, |V(r)|^{1/2}+\frac{1}{4\pi} \ln\left[\frac{V(p)}{V(q}\right]+\frac{1}{2},
\end{equation}
where $p$ and $q$ are defined by
\begin{equation}
\label{eq27b}
\int_0^p dr\, |V(r)|^{1/2}=\frac{\pi}{2} \quad \text{and} \quad \int_q^{\infty}dr\, |V(r)|^{1/2}=\frac{\pi}{2}.
\end{equation}
\end{subequations}

A very cogent upper limit on $L$ have been identified in Ref. \cite{br03b} (but the formula is known for a long time) and reads
\begin{subequations}
\label{eq28}
\begin{equation}
\label{eq28a}
L\leq L^+=\left\{\left\{ \sigma-\frac{1}{2}\right\} \right\},
\end{equation}
with
\begin{equation}
\label{eq28b}
\sigma=\max\left[r\,|V(r)|^{1/2}\right],
\end{equation}
\end{subequations}
where $\left\{\left\{ x\right\} \right\}$ stands for integer part of $x$. This upper limit is written for a purely negative potential (in general, $|V(r)|$ is replaced by $-V^-(r)$ where $V^-(r)$ is the negative part of the potential). 

Another upper limit on $L$ is obtained from the relation (\ref{eq24}) and reads
\begin{equation}
\label{eq29}
L\leq L^{++}=\left\{\left\{ N^+_0-1\right\} \right\}.
\end{equation}
This upper limit $L^{++}$ is obviously only applicable to potentials that satisfy (\ref{eq15}).
This relation (\ref{eq29}) together with the formula (\ref{eq26}) yields the following neat upper limit on the total number of bound states
\begin{equation}
\label{eq30}
N\leq \frac{1}{6}\, N^+_0 (N^+_0+1) (2N^+_0+1)< \frac{1}{3}\, (N^+_0 +1)^3.
\end{equation}

To conclude this Letter, we now compare the new upper limit on the total number of bound states with the exact result and with previously known upper limit. For this test, we just consider the Yukawa potential which, as already mentioned, satisfies the relation (\ref{eq15}):
\begin{equation}
\label{eq31}
V(r)=-g^2 (r R)^{-1} \exp(-r/R).
\end{equation}

The previously known upper limit yielding the best results for this potential has been obtained in the article \cite{br03b} from the {\it drastic} approximation $N_{\ell}\le N^+_0$, with $N^+_0$ given by (\ref{eq27}) and $L^+$ given by (\ref{eq28}). The fact that, in spite of this poor approximation, the results obtained are the more accurate clearly indicates that in general upper limits on the total number of bound states are not very cogent and yields poor results. This limit reads
\begin{equation}
\label{eq32}
N<\frac{1}{4}(2\sigma+1)^{2}\,N^+_0,
\end{equation}
where $N^+_0$ is the right-hand side of (\ref{eq27a}) and $\sigma$ is defined by (\ref{eq28b}). We can also consider the celebrated Lieb upper limit \cite{li76} as a further reference:
\begin{equation}
\label{eq33}
N<1.458\int_0^{\infty} dr\, r^2\, |V(r)|^{3/2}.
\end{equation}

Instead of presenting a detailed comparison, we will just compare the leading term of each upper limit. This is enough to show that the new upper limit is indeed more stringent. Moreover, one can verify that the new upper limit yields the strongest restriction as soon as $g$ is greater than $1.9$. For this value of the strength of the potential, there is just one S-wave bound state. Then, as soon as there is more than one bound state in the potential, the new upper limit is better. We show now that the improvement is indeed significant. The exact asymptotic behavior of the total number of bound states is given by \cite{ma72}
\begin{equation}
\label{eq34}
N\approx \frac{2}{3\pi}\int_0^{\infty} dr\, r^2\, |V(r)|^{3/2}.
\end{equation}
For the Yukawa potential, the exact asymptotic formula (\ref{eq34}) gives
\begin{equation}
\label{eq35}
N\approx\frac{2}{9}\sqrt{\frac{2}{3\pi}}\, g^3=0.102\,g^3.
\end{equation}
The Lieb formula (\ref{eq33}) yields
\begin{equation}
\label{eq36}
N\approx 0.703\,g^3.
\end{equation}
The upper limit (\ref{eq32}) leads to
\begin{equation}
\label{eq37}
N\approx 0.294 \,g^3,
\end{equation}
which is already a nice improvement compared to the Lieb formula. The new upper limit (\ref{eq26}) gives
\begin{equation}
\label{eq38}
N\approx 0.145 \,g^3.
\end{equation}
The numerical coefficient of (\ref{eq38}) is still $1.42$ times too large compared to the exact asymptotic formula (\ref{eq35}) but the improvement over the result (\ref{eq37}) is important (about a factor $2$). The neater version (\ref{eq30}) of the new upper bound gives asymptotically
\begin{equation}
\label{eq39}
N\approx 0.169 \,g^3,
\end{equation}
which is reasonably close to the result (\ref{eq38}).

\end{document}